\documentclass[sigconf]{acmart}

\setcopyright{none}
\usepackage{todonotes}
\usepackage[compact]{titlesec}
\renewcommand\footnotetextcopyrightpermission[1]{}

\AtBeginDocument{%
  \providecommand\BibTeX{{%
    \normalfont B\kern-0.5em{\scshape i\kern-0.25em b}\kern-0.8em\TeX}}}

\acmConference[CoDS-COMAD '20]{CoDS-COMAD '20: ACM India Joint International Conference on 
Data Science & Management of Data}{January 05--07, 2020}{Hyderabad, India}
\acmBooktitle{7th ACM IKDD CoDS and 25th COMAD (CoDS-COMAD '20),
  January 05--07, 2020, Hyderabad, India}
\acmPrice{15.00}
\setcopyright{acmcopyright}

\emergencystretch=\maxdimen
\begin{document}

\title{All It Takes is 20 Questions!: A Knowledge Graph Based Approach }

\author{Alvin Dey, Harsh Kumar Jain, Vikash Kumar Pandey, Tanmoy Chakraborty}
\affiliation{%
  \institution{IIIT-Delhi, India}}
\email{{alvin18066,harsh18006,vikash18086,tanmoy}@iiitd.ac.in}

\begin{abstract}
20 Questions (20Q) is a two-player game. One player is the answerer, and the other is a questioner. The answerer chooses an entity from a specified domain and does not reveal this to the other player. The questioner can ask at most 20 questions to the answerer to guess the entity. The answerer can reply to the questions asked by saying {\em yes/no/maybe}. 

In this paper, we propose a novel approach based on knowledge graph for designing the 20Q game on Bollywood movies. The system assumes the role of the questioner and asks questions to predict the movie thought by the answerer. It uses a probabilistic learning model for template-based question generation and answers prediction. A dataset of interrelated entities is represented as a weighted knowledge graph, which updates as the game progresses by asking questions. An evolutionary approach helps the model to gain a better understanding of user choices and predicts the answer in fewer questions over time. Experimental results show that our model was able to predict the correct movie in less than 10 questions for more than half of the times the game was played. This kind of models can be used to design applications that can detect diseases by asking questions based on symptoms, improving recommendation systems, etc.
\end{abstract}


\maketitle

\section{Introduction}
The 20 Questions game started as a spoken parlor game in early 19\textsuperscript{th} century. During the early phases, it was known as `Animal, Plant and Mineral'. In this version, the answerer was supposed to tell the questioner about the category he/she has chosen. The principle behind this game is that one player thinks of an entity, and the other player asks a series of (maximum 20) questions to guess that entity. These questions should be answerable in {\itshape yes/no/maybe}.

Mathematically, the game allows identifying 2\textsuperscript{20} arbitrary objects where each question eliminates half the entities. Therefore, a practical strategy would be to ask questions in such a way that reduces the list of possible answers roughly into half. This game has a huge potential in real-life applications. It can be used to develop healthcare applications, where the patient answers simple questions, and the system predicts the disease. The application models the question over symptoms like:
\begin{description}
  \item[$\bullet$] {\itshape Are you feeling cold?}
  \item[$\bullet$] {\itshape Are you feeling nauseous?}
  \item[$\bullet$] {\itshape Are you having a headache?}
\end{description}
These systems help to collect data about human health and improve healthcare facilities. In this paper, we test our model on a huge set of Bollywood movies. The reason for choosing Bollywood movies is the popularity of Bollywood and the humongous metadata available to build such a large-scale system.

The existing baseline systems try to model the same problem. However, they have not been able to develop a system which takes care of human errors, e.g., if unknowingly, the answerer answers a question incorrectly, then the system should be intelligent enough not to eliminate all the possibilities.
For example, answering {\itshape`no'}  to a question like `Was Aamir Khan an actor of your movie?' should not blatantly remove the possibility of a movie like `3 Idiots' if the answerer may have answered it incorrectly. The probability of getting `3 Idiots' may decrease, but if the answerer answers all other questions correctly, then the model is still expected to predict the correct movie.

In this paper, we present a novel approach to predict movies in 20Q game using a knowledge graph and a probabilistic learning model that evolves as the game is played and predicts correct movie in less than 20 questions.
We design the system in five individual segments (discussed in detail in model architecture). The model starts with equal probability for every movie, which changes over subsequent questions. 
It attains fault tolerance as it re-balances the movies probabilities in a way, that it does not disregard or accept a movie completely after every answer.
The question generator poses questions based on three components:
\begin{enumerate}
\item Probability from past experience.
\item Probability based on the density of edge connectivity in the knowledge graph.
\item Cumulative probability of movies under a category during the current run (based on player's responses).
\end{enumerate}

The proposed model overcomes all the existing challenges of the baseline models. The major contributions of the paper are mentioned below:

\begin{description}
  \item[$\bullet$] We collected a dataset of 18,481 Indian movies from DBpedia\footnote{\url{https://wiki.dbpedia.org/}}. The dataset includes 113 features per movie such as movie name, movie length, director, producer, actors, genre, subject, etc.
  \item[$\bullet$] We developed and evaluated a novel architecture using knowledge graph to predict Bollywood movies using 20Q game.
  \item[$\bullet$] The proposed model is robust enough to handle incorrect answers given by the answerer. It predicts the correct answers in 90.8\% of cases. In 50\% cases, it predicts correctly by asking less than 10 questions.
\end{description}
The codes and datasets are available publicly at {\color{blue}\url{https://github.com/harshj94/20Q-Game}}.

\section{Related Work}

During the initial phases of AI and NLP, the primary notion for question answering (QA) was that machines would be able to answer by converting the question to a machine-readable form and then match it against a background knowledge stack. However, no such system has been built to represent questions definitively.
START \cite{Katz2005ExternalKS} was the first QA model, based on parsing through structured data for answer prediction. 
Similar to human approach, the QA model was paired with the ability to find relations between possible questions for efficient answer set reduction.

In the AURA and HALO models \cite{halo_update}, the answers were formulated from questions based on structured documents containing principles from the related field. AURA faced issues while relating query 
templates with underlying entities.

Exiting open-ended QA systems deploy a pipeline of passage search based technique against a related corpus to generate possible answers which match the expected type.
For the search component, web information has been used for generating possible answers \cite{Clarke_webreinforced, Dumais02webquestion, integrating_web} as well as confirming present possibilities \cite{Magnini:2002:RAE:1073083.1073154, answer_selection}.
Wikipedia and other online resources have been referenced as a standard corpora by many Question Answering models \cite{Kupiec:1993:MRL:160688.160717, Ahn2004UsingWA} as well as in CLEF validation metric \cite{10.1007/978-3-540-85760-0_27}. However, these models treated the corpus mentioned above as a Newswire corpus extension. They did not use their underlying properties to improve performance.

From the answer generation viewpoint, existing QA models formulate a semantic-type based method to suggest answers which try to match the expected answer type \cite{Prager:2000:QPA:345508.345574, moldovan-etal-2000-structure}.
In contrast, our model does not depend on the nature of the problem -- it tries to gather a possible candidate set based on associated meta-tags of the movies such as director, actor, release year, etc.
Even though our method generates a much broader set of candidate answers, it can outperform semantic-based methods in a wider array of fields.

Advanced techniques in Deep learning have shown remarkable performance in QA \cite{Tay:2018:MAN:3219819.3220048, DBLP:journals/corr/abs-1905-12897, tayyar-madabushi-etal-2018-integrating}. Zheng et al. \cite{Zheng:2018:QAO:3236187.3269455} tried to find `help' documents based on the user query. The user query is used to find out the best possible query template using text mining and simultaneously building a semantic dependency graph. We design our algorithm to reduce the size of possible answers from the knowledge graph as discussed in \cite{Zheng:2018:QAO:3236187.3269455}.
Yang et al.  \cite{Yang:2017:EAT:3298483.3298686} proposed a solution for relating questions to relevant documents using a probabilistic scoring approach. Here we use a similar mechanism for questionnaire generation.

Most of the existing studies develop an answering model for the questions asked by the user. However, there are a  few models \cite{million_queries, needle_haystack}, which create carefully curated questions to predict the information as per the user's answers, like the model discussed in this paper.

\section{Dataset Details}
We use DBpedia, which is a structured content extraction project created using information from Wikipedia. It allows us to semantically query relations on properties related to Wikipedia resources and other datasets. We extract data of 18481 movies from DBpedia. Each movie has 113 metadata tags associated with it.

\subsection{Data Acquisition}
\begin{figure}[t]
\includegraphics[scale=0.6]{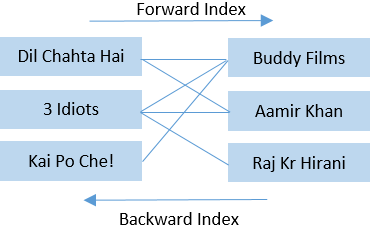}
\caption{Sample of data representing the forward and the backward indexing between movies and metadata.}
\label{fig:forw_back}
\end{figure}

\begin{enumerate}
  \item We acquire the dataset and formulate the knowledge graph using a two-way inverted index.
  \item The forward inverted index maps movies to its respective metadata tags.
  \item The backward index maps metadata tags to all the movies associated with it.
\end{enumerate}
Figure \ref{fig:forw_back} shows the forward and backward index on a subset of data.

\begin{figure*}[ht]
\includegraphics[scale=0.6]{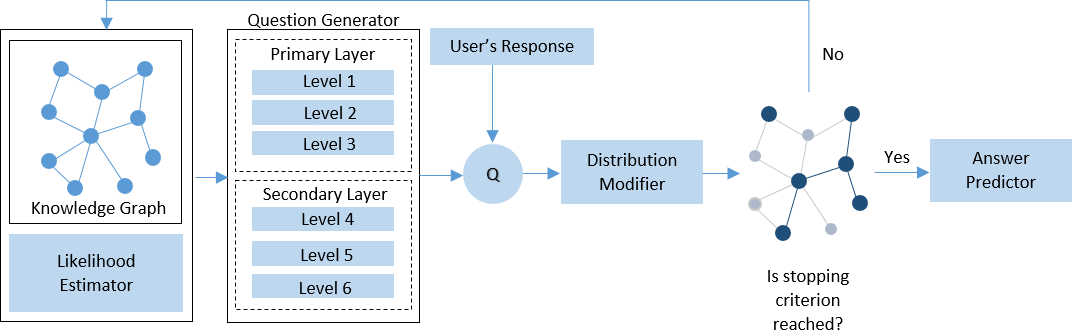}
\caption{Model architecture of the proposed system. The initial knowledge graph with equiprobable nodes along with likelihood estimator is provided to the question generator. It generates a question (Q) from one of the levels. The user's response to the question modifies the probabilities of nodes in the graph. If the stopping criteria are met, the model predicts the answer; else, the system iterates using the updated graph.}
\label{fig:model_arch}
\end{figure*}

\subsection{Data Preprocessing}
We filter the data during preprocessing by eliminating redundancies and inconsistency. The preprocessing details are as follows:
\begin{enumerate}
  \item Reduce the dataset to 200 popular movies for conducting the experiments.
  \item Remove the tags that are present in less than 10\% of the dataset.
  \item Create additional tags like `Era' to signify the decade in which the movie was released.
  \item Filter out and keep only relevant values from tags like `Genre' and `Subject', e.g., {\itshape Indian crime films, Indian romance films, Indian thriller films} and remove values like {\itshape masala films, circus films, films about courtesans in India} that are generally used less.
  \item Manually add the missing values for movies in which higher-order attributes such as Director, Music composer, etc. were missing.
\end{enumerate}

\section{Proposed Approach}
We broadly classify the questions into two layers: (i) primary layer and (ii) secondary layer. The primary layer questions are focused on a wider range of movies, while the secondary layer questions are more specific in nature and targeted towards a smaller set of movies. Figure \ref{fig:model_arch} shows the architectural details of the model.
\begin{table}[!t]
  \caption{Different primary and secondary questions generated by the system during a game play}
  \label{tab:freq}
  \begin{tabular}{p{3.8cm}|p{3.8cm}}
    \toprule
    Primary Questions&Secondary Questions\\
    \midrule
    Is your movie from the 1990s era? & Is Aamir Khan an actor of your movie?\\
    Is Bollywood romance the genre of your movie? & Is Karan Johar the director of your movie?\\
    Is feminist films the subject of your movie? & Is A.R. Rahman the music composer of the movie?\\
  \bottomrule
\end{tabular}
\vspace{-5mm}
\end{table}
Our model is divided into five components:

{\bfseries 1. Question Generator:}
The generator is a template-based hierarchically structured model. It traverses the knowledge graph to ask  questions based on -- learned experiences, the answers it received during the current run and the most likely movies based on scores assigned to each movie. The architecture poses the questions taking into account user-specific data in the primary layers to reduce the size of the most probable set. The secondary layer poses tricky questions specific to a limited set of movies, to get an in-depth insight into the choices. Table \ref{tab:freq} shows instances of primary and secondary layer questions.

{\bfseries 2. Answer Predictor:}
The predictor outputs a list of five movies in descending order of their probabilities. It makes a guess once the total probability of the top five most likely movies reaches the empirical value of 0.5. The predictor removes the movies from the probable choices if the user replies {\em no} to these five guesses. If the user says {\em yes} the game stops and asks the user for the exact movie(from the 5 movies). It then alters the edge probabilities in the graph for future games. We perform this adjustment as every choice a player makes is an indication of the popularity of the movie and it's associated entities.

{\bfseries 3. Likelihood Estimator:}
For the {\bf primary layer}, we store two different probability values: (i) probability on each level which is decided by the number of movies in which the specific entity is present, and (ii) how many times any user has elected that particular entity on the given level.

Let \( p_v(l) \) denote the probability of entity \( v \) at level \( l \), and \( p_v(h) \) denote the probability of entity \( v \) stored throughout the run of the game denoted by \( h \). We assign the total probability as weighted sum of both the probabilities:
\begin{equation}
p_v(f)=\alpha.p_v(l)+(1-\alpha).p_v(h)
\end{equation}
Here \( \alpha \) is preset to 0.2. This is to ensure that the model learns more from the games played so far rather than the static scores at each level.
We add an additional component for user specific likelihood estimation. For example. to estimate the era of a movie, we use the following formula:
\begin{equation}\small
p_v(e)=\sum_{10\,years}\left(\ln \left(\frac{1}{10\sqrt{2\pi}}\right)-\left(\frac{\text{year}-\left(\text{birthyear}\textsubscript{user}+20\right)^2}{200}\right)\right)
\end{equation}
\begin{equation}
score(l)=p_v(f) + p_v(e)
\end{equation}

For the {\bf secondary layer}, an additional probability component for the total distribution score \( d \) under entity \( v \) during the current run of the game is computed as follows:
\begin{equation}
p_v(f)=\alpha.p_v(l)+(1-\alpha).p_v(h)+\beta.p_v\left(\sum_{n\in set(v)}d(n)\right)
\end{equation}
where, set(v) denotes the set of movies under the entity  \( v \).

{\bfseries 4. Distribution Modifier:}
For cases, where the user answers {\itshape maybe}, the distribution remains unchanged. For definitive answers, the distribution is modified as follows:
\begin{equation}
M_t(i)=\frac{(M_{t-1}(i)\times e^{-\alpha_{t-1}\times-\beta{t-1}})}{Norm(t-1)}
\end{equation}
\( \beta_{t-1}=+1 \) for the set of movies where the user's answer is {\em no}, or \(-1\) for {\em yes}. \( \alpha_{t-1}\) is fixed to 1 empirically. \( M_{t-1}(i)\) is the current distribution. \( Norm(t-1)\) is the normalization factor for current distribution. 

The Modifier essentially takes the set of movies for which the user answered {\em yes} and increases their probability, while decreasing the probability of the set of movies for which the user answered {\em no} (using {\em eqn} 5). The set of predicted movies for which the user evaluates as {\em incorrect}, we distribute their probability among all the remaining movies equally and set theirs to 0. 

We perform the above step for two reasons: Distributing the probability equally won't change the relative difference in score of all the other movies and the set of movies for which the user answered {\em no} will never appear as a guess again.

{\bfseries 5. Answer Tracer:} 
Every time the model predicts the movie, the answerer is asked if the prediction is correct. The next prediction, therefore, is based on the response of the answerer. If the system is unable to predict within 20 questions, it gives a trace of user answers along with the corresponding facts related to the movie.

\section{Baseline Methods}
We compare our model with the following baselines (we designed for this task due to the unavailability of existing methods particularly for this task). We use questions asked and cumulative probability of ranked guesses as the evaluation metrics to study the comparative analysis.

\begin{description}
  \item[$\bullet$] \textbf{Baseline 1:} The model frames questions systematically from six aspects of a movie -- era, genre, subject of the story, actors, director, and music composer. The questions eliminate a subset of possible answers after a definite reply by the user. An answer as {\em maybe} does not contribute to the understanding of the model and retains the current state. The model poses questions based on the possibilities it gathers over the current run of answers. It eliminates answers in a strict binary fashion without due regard to human fallacies during the game. Figure \ref{fig:que_selection} highlights the game proceedings for question selection.
  \begin{figure}[t]
  \includegraphics[scale=0.5]{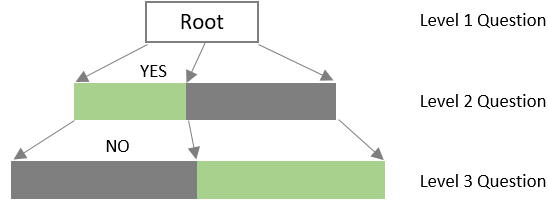}
  \vspace{-3mm}
  \caption{Question selection model in Baseline 1.}
  \label{fig:que_selection}
  \vspace{-3mm}
  \end{figure}
  \item[$\bullet$] \textbf{Baseline 2:} This model frames questions from the same six aspects of a movie as baseline 1 along with a learning model added to the graph traversal. It poses questions hierarchically, giving weight to initial questions with the maximum possibility of answer set reduction. It also takes into account the user's personal information such as `birth year' to adapt the timeline of movies to be questioned for a more efficient guess. The model associates probability scores to each subcategory of the aspects mentioned above to determine the most likely category. Like baseline 1, this model also suffers from the lack of robustness towards human errors. However, due to the learning aspect it performs more efficiently than baseline 1. 
\end{description}

\section{Experiments and Results}
The evaluation is conducted through a user case study. A total of 50 participants interacted with the game. Each participant played the game 5 times for different movies. They played with the same set of movies with each of the baselines as well. 21 participants were female, and 29 were male. Their age group distribution is shown in Table \ref{tab:age_dist}.

\begin{table}[t]
  \caption{Age group distribution of human participants. }
  \label{tab:age_dist}
  \vspace{-3mm}
  \begin{tabular}{c|c}
    \toprule
    Age group (in years)&No. of participants\\
    \midrule
    18-25&37\\
    25-35&8\\
    35+&5\\
  \bottomrule
\end{tabular}
\vspace{-5mm}
\end{table}

We trained baseline 2 and our model for 50 random movies before the participants interacted with them. This provides a preliminary idea about the distribution to the model so that it won't suffer from the problem of a cold start. In each game, every competing model generated 5 movies as output in a ranked fashion. Each output is considered as a single attempt for the evaluation.

\begin{table}[]

  \caption{Number of questions asked to predict the correct movie. The game was played 250 times.}
  \label{tab:noq}
  \vspace{-3mm}
  \scalebox{0.7}{
  \begin{tabular}{c|c|c|c}
    \toprule
    No. of questions&Baseline 1&Baseline 2&Proposed model\\
    \midrule
    <10     &       37      &   72  &   127\\
    10-15   &       38      &   93  &   62\\
    15-20   &       109     &   39  &   38\\
    Not Answered&   66      &   46  &   23\\
  \bottomrule
\end{tabular}}
\vspace{-5mm}
\end{table}

\begin{table}[t]
  \caption{Cumulative probability of the competing models to predict the correct movie within $n$\textsuperscript{th} rank (1<=$n$<=5).}
  \label{tab:prob}
  \vspace{-3mm}
  \begin{tabular}{c|c|c|c}
    \toprule
    Rank&Baseline 1&Baseline 2&Proposed model\\
    \midrule
    1   &   0.032   &   0.104   &   0.564\\
    2   &   0.052   &   0.148   &   0.712\\
    3   &   0.064   &   0.196   &   0.788\\
    4   &   0.096   &   0.232   &   0.828\\
    5   &   0.108   &   0.256   &   0.860\\
  \bottomrule
\end{tabular}
\vspace{-5mm}
\end{table}

Table \ref{tab:noq} shows the comparison of the different number of questions that were asked before a model predicted the correct movie. Baseline 2 shows significant improvement over Baseline 1 because of the learning aspect incorporated into the model. Baseline 2 poses relevant questions during the initial stages using previous experiences. However, our model outperforms the two baselines as it predicts the correct movie within 10 questions for \textasciitilde50\% of the times while maintaining an error of \textasciitilde10\%, for which the model was unable to guess the movie within 20 questions.

Table \ref{tab:prob} shows the cumulative probability of the ranks within which the movie is predicted correctly in the first attempt. Our model significantly outperforms both the baselines. 

Our model ranks the movies effectively in lesser number of question as it learns to assign probabilities to each movie rather than just associating them with metadata.
It predicts the correct movie even when the participants answer few questions incorrectly. This is evident from the number of movies that each model was not able to predict (Table \ref{tab:noq}). Answering initials questions incorrectly makes it tough for our model to predict the movie within one attempt. However, it recovers and predicts the movie in later stages of the game in most of the cases.

\section{Conclusion}

An essential aspect for achieving good QA accuracy is to make sure the correct answer remains in the candidate set. It needs to be maintained despite human errors in judgment. Taking this into account, we propose a knowledge graph-based approach to develop a 20Q game on Bollywood data. The model overcomes the major issue present in the baselines of handling human errors in answering questions by distributing probabilities intelligently. Our model predicted correct movies in fewer questions as compared to the baselines.

This work can be extended to improve recommendation systems, create an application that can ask simple questions and predict class, etc. The question generator can be further extended to generate questions based on the context of the movie plot, characters, script, etc. using Sequence-to-Sequence model. It helps the model to ask more specific questions and predict correct values effectively.
\bibliographystyle{ACM-Reference-Format}
\bibliography{knowledge-graph}


\begin{thebibliography}{19}


\ifx \showCODEN    \undefined \def \showCODEN     #1{\unskip}     \fi
\ifx \showDOI      \undefined \def \showDOI       #1{#1}\fi
\ifx \showISBNx    \undefined \def \showISBNx     #1{\unskip}     \fi
\ifx \showISBNxiii \undefined \def \showISBNxiii  #1{\unskip}     \fi
\ifx \showISSN     \undefined \def \showISSN      #1{\unskip}     \fi
\ifx \showLCCN     \undefined \def \showLCCN      #1{\unskip}     \fi
\ifx \shownote     \undefined \def \shownote      #1{#1}          \fi
\ifx \showarticletitle \undefined \def \showarticletitle #1{#1}   \fi
\ifx \showURL      \undefined \def \showURL       {\relax}        \fi
\providecommand\bibfield[2]{#2}
\providecommand\bibinfo[2]{#2}
\providecommand\natexlab[1]{#1}
\providecommand\showeprint[2][]{arXiv:#2}

\bibitem[\protect\citeauthoryear{Ahn, Jijkoun, Mishne, Müller, de~Rijke, and
  Schlobach}{Ahn et~al\mbox{.}}{2004}]%
        {Ahn2004UsingWA}
\bibfield{author}{\bibinfo{person}{David Ahn}, \bibinfo{person}{Valentin
  Jijkoun}, \bibinfo{person}{Gilad Mishne}, \bibinfo{person}{Karin Müller},
  \bibinfo{person}{Maarten de Rijke}, {and} \bibinfo{person}{Stefan
  Schlobach}.} \bibinfo{year}{2004}\natexlab{}.
\newblock \showarticletitle{Using Wikipedia at the TREC QA Track.}, In
  \bibinfo{booktitle}{TREC}.
\newblock \bibinfo{journal}{\emph{Journal of Colloid and Interface Science - J
  COLLOID INTERFACE SCI}}.
\newblock


\bibitem[\protect\citeauthoryear{Chu-Carroll, Fan, Boguraev, Carmel, Sheinwald,
  and Welty}{Chu-Carroll et~al\mbox{.}}{2012}]%
        {needle_haystack}
\bibfield{author}{\bibinfo{person}{Jennifer Chu-Carroll}, \bibinfo{person}{J
  Fan}, \bibinfo{person}{B.K. Boguraev}, \bibinfo{person}{David Carmel},
  \bibinfo{person}{Dafna Sheinwald}, {and} \bibinfo{person}{Christopher
  Welty}.} \bibinfo{year}{2012}\natexlab{}.
\newblock \showarticletitle{Finding needles in the haystack: Search and
  candidate generation}.
\newblock \bibinfo{journal}{\emph{IBM Journal of Research and Development}}
  \bibinfo{volume}{56} (\bibinfo{date}{05} \bibinfo{year}{2012}),
  \bibinfo{pages}{6:1--6:12}.
\newblock
\urldef\tempurl%
\url{https://doi.org/10.1147/JRD.2012.2186682}
\showDOI{\tempurl}


\bibitem[\protect\citeauthoryear{Clarke, Cormack, Lynam, Li, and
  McLearn}{Clarke et~al\mbox{.}}{2001}]%
        {Clarke_webreinforced}
\bibfield{author}{\bibinfo{person}{C.~L.~A. Clarke}, \bibinfo{person}{G.V.
  Cormack}, \bibinfo{person}{T.R. Lynam}, \bibinfo{person}{C.M. Li}, {and}
  \bibinfo{person}{G.L. McLearn}.} \bibinfo{year}{2001}\natexlab{}.
\newblock \bibinfo{title}{Web Reinforced Question Answering (MultiText
  Experiments for TREC 2001)}.
\newblock
\newblock


\bibitem[\protect\citeauthoryear{Cohen, Amitay, and Carmel}{Cohen
  et~al\mbox{.}}{2007}]%
        {million_queries}
\bibfield{author}{\bibinfo{person}{Doron Cohen}, \bibinfo{person}{Einat
  Amitay}, {and} \bibinfo{person}{David Carmel}.}
  \bibinfo{year}{2007}\natexlab{}.
\newblock \showarticletitle{Lucene and Juru at TREC 2007: 1-Million Queries
  Track.}
\newblock


\bibitem[\protect\citeauthoryear{Dumais, Banko, Brill, Lin, and Ng}{Dumais
  et~al\mbox{.}}{2002}]%
        {Dumais02webquestion}
\bibfield{author}{\bibinfo{person}{Susan Dumais}, \bibinfo{person}{Michele
  Banko}, \bibinfo{person}{Eric Brill}, \bibinfo{person}{Jimmy Lin}, {and}
  \bibinfo{person}{Andrew Ng}.} \bibinfo{year}{2002}\natexlab{}.
\newblock \showarticletitle{Web Question Answering: Is More Always Better?}. In
  \bibinfo{booktitle}{\emph{In Proceedings of the 25th annual international ACM
  SIGIR conference on research and development in information retrieval}}.
  \bibinfo{pages}{291--298}.
\newblock


\bibitem[\protect\citeauthoryear{Giampiccolo, Forner, Herrera, Pe{\~{n}}as,
  Ayache, Forascu, Jijkoun, Osenova, Rocha, Sacaleanu, and
  Sutcliffe}{Giampiccolo et~al\mbox{.}}{2008}]%
        {10.1007/978-3-540-85760-0_27}
\bibfield{author}{\bibinfo{person}{Danilo Giampiccolo}, \bibinfo{person}{Pamela
  Forner}, \bibinfo{person}{Jes{\'u}s Herrera}, \bibinfo{person}{Anselmo
  Pe{\~{n}}as}, \bibinfo{person}{Christelle Ayache}, \bibinfo{person}{Corina
  Forascu}, \bibinfo{person}{Valentin Jijkoun}, \bibinfo{person}{Petya
  Osenova}, \bibinfo{person}{Paulo Rocha}, \bibinfo{person}{Bogdan Sacaleanu},
  {and} \bibinfo{person}{Richard Sutcliffe}.} \bibinfo{year}{2008}\natexlab{}.
\newblock \showarticletitle{Overview of the CLEF 2007 Multilingual Question
  Answering Track}. In \bibinfo{booktitle}{\emph{Advances in Multilingual and
  Multimodal Information Retrieval}}, \bibfield{editor}{\bibinfo{person}{Carol
  Peters}, \bibinfo{person}{Valentin Jijkoun}, \bibinfo{person}{Thomas Mandl},
  \bibinfo{person}{Henning M{\"u}ller}, \bibinfo{person}{Douglas~W. Oard},
  \bibinfo{person}{Anselmo Pe{\~{n}}as}, \bibinfo{person}{Vivien Petras}, {and}
  \bibinfo{person}{Diana Santos}} (Eds.). \bibinfo{publisher}{Springer Berlin
  Heidelberg}, \bibinfo{address}{Berlin, Heidelberg},
  \bibinfo{pages}{200--236}.
\newblock
\showISBNx{978-3-540-85760-0}


\bibitem[\protect\citeauthoryear{Gunning, Chaudhri, Clark, Barker, Chaw,
  Greaves, Grosof, Leung, McDonald, Mishra, Pacheco, Porter, Spaulding, Tecuci,
  and Tien}{Gunning et~al\mbox{.}}{2010}]%
        {halo_update}
\bibfield{author}{\bibinfo{person}{David Gunning}, \bibinfo{person}{Vinay~K.
  Chaudhri}, \bibinfo{person}{Peter~E. Clark}, \bibinfo{person}{Ken Barker},
  \bibinfo{person}{Shaw-Yi Chaw}, \bibinfo{person}{Mark Greaves},
  \bibinfo{person}{Benjamin Grosof}, \bibinfo{person}{Alice Leung},
  \bibinfo{person}{David~D. McDonald}, \bibinfo{person}{Sunil Mishra},
  \bibinfo{person}{John Pacheco}, \bibinfo{person}{Bruce Porter},
  \bibinfo{person}{Aaron Spaulding}, \bibinfo{person}{Dan Tecuci}, {and}
  \bibinfo{person}{Jing Tien}.} \bibinfo{year}{2010}\natexlab{}.
\newblock \showarticletitle{Project Halo Update—Progress Toward Digital
  Aristotle}.
\newblock \bibinfo{journal}{\emph{AI Magazine}} \bibinfo{volume}{31},
  \bibinfo{number}{3} (\bibinfo{date}{Jul.} \bibinfo{year}{2010}),
  \bibinfo{pages}{33--58}.
\newblock
\urldef\tempurl%
\url{https://doi.org/10.1609/aimag.v31i3.2302}
\showDOI{\tempurl}


\bibitem[\protect\citeauthoryear{Katz, Lin, Loreto, Hildebrandt, W.~Bilotti,
  Felshin, Fernandes, Marton, and Mora}{Katz et~al\mbox{.}}{2003}]%
        {integrating_web}
\bibfield{author}{\bibinfo{person}{Boris Katz}, \bibinfo{person}{Jimmy Lin},
  \bibinfo{person}{Daniel Loreto}, \bibinfo{person}{Wesley Hildebrandt},
  \bibinfo{person}{Matthew W.~Bilotti}, \bibinfo{person}{Sue Felshin},
  \bibinfo{person}{Aaron Fernandes}, \bibinfo{person}{Gregory Marton}, {and}
  \bibinfo{person}{Federico Mora}.} \bibinfo{year}{2003}\natexlab{}.
\newblock \showarticletitle{Integrating Web-based and Corpus-based Techniques
  for Question Answering.} \bibinfo{pages}{426--435}.
\newblock


\bibitem[\protect\citeauthoryear{Katz, Marton, Borchardt, Brownell, Felshin,
  Loreto, Louis-Rosenberg, Lu, Mora, Stiller, Uzuner, and Wilcox}{Katz
  et~al\mbox{.}}{2005}]%
        {Katz2005ExternalKS}
\bibfield{author}{\bibinfo{person}{Boris Katz}, \bibinfo{person}{Gregory
  Marton}, \bibinfo{person}{Gary~C. Borchardt}, \bibinfo{person}{Alexis
  Brownell}, \bibinfo{person}{Sue Felshin}, \bibinfo{person}{Daniel Loreto},
  \bibinfo{person}{Jesse Louis-Rosenberg}, \bibinfo{person}{Ben Lu},
  \bibinfo{person}{Federico Mora}, \bibinfo{person}{Stephan Stiller},
  \bibinfo{person}{{\"O}zlem Uzuner}, {and} \bibinfo{person}{Angela Wilcox}.}
  \bibinfo{year}{2005}\natexlab{}.
\newblock \showarticletitle{External Knowledge Sources for Question Answering}.
  In \bibinfo{booktitle}{\emph{TREC}}.
\newblock


\bibitem[\protect\citeauthoryear{Ko, Si, and Nyberg}{Ko et~al\mbox{.}}{2007}]%
        {answer_selection}
\bibfield{author}{\bibinfo{person}{Jeongwoo Ko}, \bibinfo{person}{Luo Si},
  {and} \bibinfo{person}{Eric Nyberg}.} \bibinfo{year}{2007}\natexlab{}.
\newblock \showarticletitle{A Probabilistic Framework for Answer Selection in
  Question Answering}. \bibinfo{pages}{524--531}.
\newblock


\bibitem[\protect\citeauthoryear{Kupiec}{Kupiec}{1993}]%
        {Kupiec:1993:MRL:160688.160717}
\bibfield{author}{\bibinfo{person}{Julian Kupiec}.}
  \bibinfo{year}{1993}\natexlab{}.
\newblock \showarticletitle{MURAX: A Robust Linguistic Approach for Question
  Answering Using an On-line Encyclopedia}. In
  \bibinfo{booktitle}{\emph{Proceedings of the 16th Annual International ACM
  SIGIR Conference on Research and Development in Information Retrieval}}
  \emph{(\bibinfo{series}{SIGIR '93})}. \bibinfo{publisher}{ACM},
  \bibinfo{address}{New York, NY, USA}, \bibinfo{pages}{181--190}.
\newblock
\showISBNx{0-89791-605-0}
\urldef\tempurl%
\url{https://doi.org/10.1145/160688.160717}
\showDOI{\tempurl}


\bibitem[\protect\citeauthoryear{Magnini, Negri, Prevete, and Tanev}{Magnini
  et~al\mbox{.}}{2002}]%
        {Magnini:2002:RAE:1073083.1073154}
\bibfield{author}{\bibinfo{person}{Bernardo Magnini}, \bibinfo{person}{Matteo
  Negri}, \bibinfo{person}{Roberto Prevete}, {and} \bibinfo{person}{Hristo
  Tanev}.} \bibinfo{year}{2002}\natexlab{}.
\newblock \showarticletitle{Is It the Right Answer?: Exploiting Web Redundancy
  for Answer Validation}. In \bibinfo{booktitle}{\emph{Proceedings of the 40th
  Annual Meeting on Association for Computational Linguistics}}
  \emph{(\bibinfo{series}{ACL '02})}. \bibinfo{publisher}{Association for
  Computational Linguistics}, \bibinfo{address}{Stroudsburg, PA, USA},
  \bibinfo{pages}{425--432}.
\newblock
\urldef\tempurl%
\url{https://doi.org/10.3115/1073083.1073154}
\showDOI{\tempurl}


\bibitem[\protect\citeauthoryear{Moldovan, Harabagiu, Pasca, Mihalcea, Girju,
  Goodrum, and Rus}{Moldovan et~al\mbox{.}}{2000}]%
        {moldovan-etal-2000-structure}
\bibfield{author}{\bibinfo{person}{Dan Moldovan}, \bibinfo{person}{Sanda
  Harabagiu}, \bibinfo{person}{Marius Pasca}, \bibinfo{person}{Rada Mihalcea},
  \bibinfo{person}{Roxana Girju}, \bibinfo{person}{Richard Goodrum}, {and}
  \bibinfo{person}{Vasile Rus}.} \bibinfo{year}{2000}\natexlab{}.
\newblock \showarticletitle{The Structure and Performance of an Open-Domain
  Question Answering System}. In \bibinfo{booktitle}{\emph{Proceedings of the
  38th Annual Meeting of the Association for Computational Linguistics}}.
  \bibinfo{publisher}{Association for Computational Linguistics},
  \bibinfo{address}{Hong Kong}, \bibinfo{pages}{563--570}.
\newblock
\urldef\tempurl%
\url{https://doi.org/10.3115/1075218.1075289}
\showDOI{\tempurl}


\bibitem[\protect\citeauthoryear{Prager, Brown, Coden, and Radev}{Prager
  et~al\mbox{.}}{2000}]%
        {Prager:2000:QPA:345508.345574}
\bibfield{author}{\bibinfo{person}{John Prager}, \bibinfo{person}{Eric Brown},
  \bibinfo{person}{Anni Coden}, {and} \bibinfo{person}{Dragomir Radev}.}
  \bibinfo{year}{2000}\natexlab{}.
\newblock \showarticletitle{Question-answering by Predictive Annotation}. In
  \bibinfo{booktitle}{\emph{Proceedings of the 23rd Annual International ACM
  SIGIR Conference on Research and Development in Information Retrieval}}
  \emph{(\bibinfo{series}{SIGIR '00})}. \bibinfo{publisher}{ACM},
  \bibinfo{address}{New York, NY, USA}, \bibinfo{pages}{184--191}.
\newblock
\showISBNx{1-58113-226-3}
\urldef\tempurl%
\url{https://doi.org/10.1145/345508.345574}
\showDOI{\tempurl}


\bibitem[\protect\citeauthoryear{Tay, Tuan, and Hui}{Tay et~al\mbox{.}}{2018}]%
        {Tay:2018:MAN:3219819.3220048}
\bibfield{author}{\bibinfo{person}{Yi Tay}, \bibinfo{person}{Luu~Anh Tuan},
  {and} \bibinfo{person}{Siu~Cheung Hui}.} \bibinfo{year}{2018}\natexlab{}.
\newblock \showarticletitle{Multi-Cast Attention Networks}. In
  \bibinfo{booktitle}{\emph{Proceedings of the 24th ACM SIGKDD International
  Conference on Knowledge Discovery \&\#38; Data Mining}}
  \emph{(\bibinfo{series}{KDD '18})}. \bibinfo{pages}{2299--2308}.
\newblock
\showISBNx{978-1-4503-5552-0}


\bibitem[\protect\citeauthoryear{Tayyar~Madabushi, Lee, and
  Barnden}{Tayyar~Madabushi et~al\mbox{.}}{2018}]%
        {tayyar-madabushi-etal-2018-integrating}
\bibfield{author}{\bibinfo{person}{Harish Tayyar~Madabushi},
  \bibinfo{person}{Mark Lee}, {and} \bibinfo{person}{John Barnden}.}
  \bibinfo{year}{2018}\natexlab{}.
\newblock \showarticletitle{Integrating Question Classification and Deep
  Learning for improved Answer Selection}. In
  \bibinfo{booktitle}{\emph{Proceedings of the 27th International Conference on
  Computational Linguistics}}. \bibinfo{publisher}{Association for
  Computational Linguistics}, \bibinfo{address}{Santa Fe, New Mexico, USA}.
\newblock


\bibitem[\protect\citeauthoryear{Yang, Zou, Wang, Yan, and Wen}{Yang
  et~al\mbox{.}}{2017}]%
        {Yang:2017:EAT:3298483.3298686}
\bibfield{author}{\bibinfo{person}{Shuo Yang}, \bibinfo{person}{Lei Zou},
  \bibinfo{person}{Zhongyuan Wang}, \bibinfo{person}{Jun Yan}, {and}
  \bibinfo{person}{Ji-Rong Wen}.} \bibinfo{year}{2017}\natexlab{}.
\newblock \showarticletitle{Efficiently Answering Technical Questions --- a
  Knowledge Graph Approach}. In \bibinfo{booktitle}{\emph{Proceedings of the
  Thirty-First AAAI Conference on Artificial Intelligence}}
  \emph{(\bibinfo{series}{AAAI'17})}. \bibinfo{pages}{3111--3118}.
\newblock


\bibitem[\protect\citeauthoryear{Yoon, Dernoncourt, Kim, Bui, and Jung}{Yoon
  et~al\mbox{.}}{2019}]%
        {DBLP:journals/corr/abs-1905-12897}
\bibfield{author}{\bibinfo{person}{Seunghyun Yoon}, \bibinfo{person}{Franck
  Dernoncourt}, \bibinfo{person}{Doo~Soon Kim}, \bibinfo{person}{Trung Bui},
  {and} \bibinfo{person}{Kyomin Jung}.} \bibinfo{year}{2019}\natexlab{}.
\newblock \showarticletitle{A Compare-Aggregate Model with Latent Clustering
  for Answer Selection}.
\newblock \bibinfo{journal}{\emph{CoRR}}  \bibinfo{volume}{abs/1905.12897}
  (\bibinfo{year}{2019}).
\newblock
\urldef\tempurl%
\url{http://arxiv.org/abs/1905.12897}
\showURL{%
\tempurl}


\bibitem[\protect\citeauthoryear{Zheng, Yu, Zou, and Cheng}{Zheng
  et~al\mbox{.}}{2018}]%
        {Zheng:2018:QAO:3236187.3269455}
\bibfield{author}{\bibinfo{person}{Weiguo Zheng}, \bibinfo{person}{Jeffrey~Xu
  Yu}, \bibinfo{person}{Lei Zou}, {and} \bibinfo{person}{Hong Cheng}.}
  \bibinfo{year}{2018}\natexlab{}.
\newblock \showarticletitle{Question Answering over Knowledge Graphs: Question
  Understanding via Template Decomposition}.
\newblock \bibinfo{journal}{\emph{Proc. VLDB Endow.}} \bibinfo{volume}{11},
  \bibinfo{number}{11} (\bibinfo{date}{July} \bibinfo{year}{2018}),
  \bibinfo{pages}{1373--1386}.
\newblock
\showISSN{2150-8097}


\end{thebibliography}

\end{document}